\documentclass[aps,prl,superscriptaddress,amsmath,amssymb,twocolumn,showpacs,floatfix,reprint]{revtex4-2}

\usepackage{braket}
\usepackage{subfigure}
\usepackage{color}
\usepackage{verbatim}
\usepackage{amsthm,amsmath,amssymb}
\usepackage{mathrsfs}
\usepackage{graphicx}
\usepackage{esint}
\usepackage{mathrsfs}
\usepackage{bm}
\usepackage{hyperref}
\usepackage{mathrsfs}
\usepackage{times}
\usepackage{soul}
\usepackage[normalem]{ulem} 

\newenvironment{localgraphicspath}[1]{
  \graphicspath{#1}
}{}

\definecolor{OliveGreen}{cmyk}{0.64, 0, 0.95, 0.40}
\definecolor{purple}{rgb}{0.6,0,0.5}

\begin{document}

\title{Quantum anomalous Hall effects and emergent $\rm{SU}(2)$ Hall ferromagnets  at fractional filling of helical trilayer graphene}

\author{Sen Niu}\email{sen.niu@csun.edu}
\affiliation{Department of Physics and Astronomy, California State University Northridge, California 91330, USA}

\author{Jason Alicea}\email{aliceaj@caltech.edu}
\affiliation{Institute of Quantum Information and Matter and Department of Physics, California Institute of Technology, Pasadena, CA 91125, USA}

\author{D. N. Sheng}\email{donna.sheng@csun.edu}
\affiliation{Department of Physics and Astronomy, California State University Northridge, California 91330, USA}

\author{Yang Peng}\email{yang.peng@csun.edu}
\affiliation{Department of Physics and Astronomy, California State University Northridge, California 91330, USA}
\affiliation{Institute of Quantum Information and Matter and Department of Physics, California Institute of Technology, Pasadena, CA 91125, USA}

\begin{abstract}
Helical trilayer graphene realizes a versatile moir\'e system for exploring correlated topological states emerging from high Chern bands. Motivated by recent experimental observations of anomalous Hall effects at fractional fillings of magic-angle helical trilayers, we focus on the higher   Chern number $|C_{band}|=2$ band and explore gapped many-body Hall states beyond the conventional Landau level paradigm. Through extensive exact diagonalization, we predict novel phases unattainable in a single $|C_{band}|=1$ band. At filling $\nu=2/3$ and $\nu=1/3$, a $\sqrt{3}\times \sqrt{3}$ charge-ordered quantum Hall crystal and a Halperin fractional Chern insulator with Hall conductance $|\sigma_{H}|=2e^2/3h$ are predicted respectively, indicating strong particle-hole asymmetry of the system. At half-filling $\nu=1/2$, an extensively degenerate  pseudospin Hall  ferromagnet featuring emergent $\rm{SU}(2)$ symmetry is found without the band being flat. Inspired by striking robustness of the ferromagnetic degeneracy, we develop a method to unveil and quantify the emergent symmetry via pseudospin operator construction  in the presence of band dispersion and Coulomb interaction, and  demonstrate persistence of the  $\rm{SU}(2)$ quantum numbers even far away from the chiral limit.  Incorporating spin-valley degrees of freedom, we identify an optimal filling regime $\nu_{\rm{total}}=3+\nu$ for realizing the above states. Notably, inter-flavor interactions renormalize the bandwidth and stabilize all the gapped phases even in realistic sublattice corrugation parameter regimes.


\end{abstract}

\date{\today }

\maketitle

{\bf \emph{Introduction.}}~Moir\'e superlattices, engineered to host flat topological electronic bands with dramatically enhanced electron-electron interactions, have emerged as a versatile platform to explore correlated topological phases~\cite{andrei2021marvels,mak2022semiconductor}. One of the major breakthroughs in this field is the observation of zero-field fractional Chern insulators~\cite{tang2011high,sun2011nearly,neupert2011fractional,sheng2011fractional,regnault2011fractional} (FCIs) in twisted transition metal dichalcogenides (TMDs) \cite{cai2023signatures,zeng2023thermodynamic,park2023observation,xu2023observation} and rhombohedral pentalayer graphene/hBN moir\'e superlattices \cite{lu2024fractional}, paving a new route toward anyon-based quantum devices.  The phenomenology has been enriched further still by 
the discovery of quantum Hall crystals (QHCs) which exhibit quantized Hall conductance alongside charge density wave (CDW) order~\cite{tevsanovic1989hall,pan2022topological,xu2024maximally,song2024intertwined,sheng2024quantum,dong2024theory,dong2024anomalous,dong2024stability,soejima2024anomalous,tan2024parent,paul2024designing,patri2024extended} in various graphene multilayers~\cite{polshyn2022topological,waters2024interplay,lu2025extended,su2025moire}.

Moir\'e systems feature exquisitely tunable electronic structure via changes in twist angle and number of layers.
Helical trilayer graphene (HTG), composed of three layers of graphene with successive twist angles $\theta$, displays a super-moir\'e structure on length scale $1/\theta^2$ forming symmetry-related domains~\cite{hoke2024imaging} as well as a finer moir\'e pattern on a $1/\theta$ length scale~\cite{nakatsuji2023multiscale,devakul2023magic,guerci2024chern,guerci2024nature,yang2024multi}. Within individual domains, the system hosts degenerate $C_{band}=\pm1,\mp2$ Chern bands separated from remote bands~\cite{devakul2023magic,guerci2024chern,guerci2024nature}. Experimental observations of anomalous Hall effects at integer fillings  confirm the existence of these Chern bands, yet the nature of correlated states observed at fractional fillings $\nu_{\rm total}=2/3,7/2$ remains unresolved~\cite{xia2025topological}. 

Recent theoretical studies have also proposed mechanisms for correlated states in HTG~\cite{kwan2024strong,kwan2024fractional}, including integer filling states and $\nu=1/3$ Laughlin-type FCIs stabilized in an isolated $|C_{band}|=1$ band. 
However, the more captivating $|C_{band}|=2$ band in HTG remains unexplored, which could host exotic multicomponent Hall states beyond those accessible in a $|C_{band}|=1$ band~\cite{barkeshli2012topological, wu2014haldane, liu2021gate, wang2023origin, wilhelm2023non, dong2023many}. In the chiral limit which  emulates the Landau level physics, it is known that high Chern bands can support Halperin-type FCIs and quantum Hall pseudospin ferromagnets (QHFs)~\cite{dong2023many}. 
However, so far the understanding of the exotic QHF is very primitive. The state hinges on an emergent $\rm{SU}(2)$ structure and has thus far only been constructed in the chiral limit, so its mechanism and stability under non-chiral conditions remain open issues. Moreover, in actual materials, the emergence of states like FCIs are typically challenged by band dispersion and imperfect quantum geometry~\cite{xie2021fractional}.
Therefore, exploring correlated Hall states at fractional filling and searching for optimal realistic parameter regimes to stabilize them  becomes crucial. A comprehensive understanding of the many-body phase diagram and stability may further elucidate the nature of the experimentally observed $\nu=1/2, 2/3$ states and shed light on why a $\nu=1/3$ state is not yet observed.

\begin{figure}[hbt!]
\begin{localgraphicspath}{{figs/draft_fig/}}
\includegraphics[width=1\columnwidth]{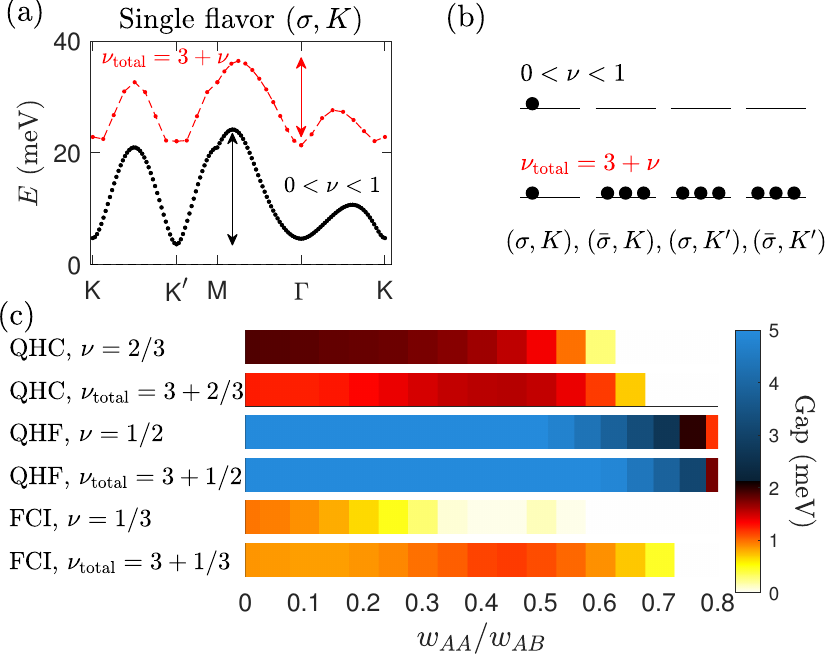}
\caption{{\bf Chern bands and qualitative many-body phase diagrams.} (a) Single-particle Chern bands (black curve) with $C_{band}=-2$ and mean-field renormalized dispersion (red curve) for $w_{AA}/w_{AB}=0.6$. (b) Schematic flavor polarization at two filling regimes.  (c) Phase diagrams for quantum Hall crystal (QHC), quantum Hall ferromagnet (QHF) and fractional Chern insulator (FCI) states at filling regimes $0<\nu<1$  and $\nu_{\rm{total}}=3+\nu$. Results for $\nu=2/3$, $1/2$, $1/3$ are obtained on $N_s=27,28,24$ clusters, respectively. }
\label{fig:band}
\end{localgraphicspath}
\end{figure}

In this letter, we investigate correlated states in the higher Chern band of HTG. Close to the magic angle, we show that the 
$|C_{band}|=2$ band can be energetically decoupled from other bands [Fig.~\ref{fig:band} (a)] using 
substrate-induced sublattice potentials. 
Through extensive exact diagonalization, we uncover a rich phase diagram including a $\sqrt{3}\times\sqrt{3}$ QHC at $\nu=2/3$, a Halperin-type FCI at $\nu=1/3$, and a QHF at $\nu=1/2$ with pseudospin ferromagnetsm. Among the three states, the QHF is particularly interesting due to its large spectral gap and its emergent $\rm{SU}(2)$ pseudospin symmetry. We unravel the $\rm{SU}(2)$ structure via the construction of  spin operators and show that it holds well even for dispersive bands. While the single-particle kinetic energy competes with the Coulomb interaction, inter-flavor (spin/valley) interactions induce a bandwidth renormalization at filling $\nu_{\rm{total}}=3+\nu$ [Fig.~\ref{fig:band} (b)]. 
The effectively enhanced interactions stabilize all the gapped many-body states at realistic parameter regimes [Fig.~\ref{fig:band} (c)]. 

{\bf \emph{Higher Chern band and many-body Hamiltonian.}}~The helical trilayer Hamiltonian for a single flavor (spin and valley) takes the form~\cite{devakul2023magic,guerci2024chern,guerci2024nature}
\begin{align}
H_{0}=\left(\begin{array}{ccc}
\hbar v_{0}\mathbf{\hat{k}}\cdot\boldsymbol{\sigma} & T(\bold r,\boldsymbol{\Phi}) & 0\\
h.c. & \hbar v_{0}\mathbf{\hat{k}}\cdot\boldsymbol{\sigma} & T(\bold r,-\boldsymbol{\Phi})\\
0 & h.c. & \hbar v_{0}\mathbf{\hat{k}}\cdot\boldsymbol{\sigma}
\end{array}\right),
\label{eq:H0} 
\end{align}
where $\sigma_{x,y,z}$ act on A/B sublattices, $\mathbf{\hat{k}}=(V k_x, k_y-VK_y^j)$ is layer-dependent momentum center with $V=\mp 1$ labeling the $K,K'$ valleys, and $K_y^j=(j-2)k_{\theta}$ represents the momentum shifts for three layers $j=1,2,3$. 
Throughout we assume Fermi velocity $v_0=10^6$m/s, moir\'e lattice spacing $a_M=a_0/2\sin{\frac{\theta}{2}}$ with graphene lattice constant $a_0 = 2.46$\AA, and reciprocal wavevector $k_{\theta}=4\pi/3a_M$. The tunneling potential $T(\bold r,\boldsymbol{\Phi})$ for $K$ valley [$T(\bold r,\boldsymbol{\Phi})^\dagger$ for $K'$ valley] takes the form
$T(\bold r,\boldsymbol{\Phi})=\sum_{n}T_{n}e^{-i\Phi_{n}}e^{-i\mathbf{q}_{n}\cdot \mathbf{r}}$, 
where the tunneling matrices $T_{n}$ are defined as $T_{n}=w_{AA}\sigma_0 +w_{AB}[\sigma_x \cos{\frac{2\pi(n-1)}{3}}  -\sigma_y \sin{\frac{2\pi(n-1)}{3}}]$ with $w_{AB}=110$ meV. The momentum vectors are $\mathbf{q}_{1}=k_{\theta}(0,-1),\mathbf{q}_{2}=k_{\theta}(\sqrt{3}/2,1/2),\mathbf{q}_{3}=k_{\theta}(-\sqrt{3}/2,1/2)$, and the inter-layer offset $\boldsymbol{\Phi}=\frac{2\pi}{3}(0,1,-1)$ is fixed as a constant inside a domain~\cite{devakul2023magic,guerci2024chern}. 

In the chiral limit ($\omega_{AA}=0.0$) and at the magic angle $\theta=1.5^{\circ}$, HTG exhibits two exactly flat, degenerate central bands carrying a total Chern number $-1$ for the $K$ valley ($+1$ for $K'$). When expressed in the sublattice-polarized basis, these split into one band that mimics a single Landau level with $C_{band}=1$ and another higher Chern band with $C_{band}=-2$~\cite{devakul2023magic,guerci2024chern}. To energetically separate the two bands, we consider an external sublattice potential $H_{\boldsymbol{\mu}}=\text{diag}(\mu_{1}\sigma_z,\mu_{2}\sigma_z,\mu_{3}\sigma_z)$ which can be induced by alignment to a hexagonal boron nitride (hBN) substrate~\cite{giovannetti2007substrate}. In the idealized case $\mu_1=\mu_2=\mu_3$, the two bands split rigidly without modifying their dispersions. Here we choose $\boldsymbol{\mu}=(-4,-6,-8)$ meV, reflecting a decaying hBN-induced potential from bottom to top layer. For this study, we fix the twist angle at $\theta=1.44^{\circ}$ (near the magic angle) and use $w_{AA}=0.6 w_{AB}$ to model realistic corrugation. Due to the $\boldsymbol{\mu}$ potential, the non-interacting $C_{band}=-2$ band acquires positive kinetic energy as shown in Fig.~\ref{fig:band} (a), separated from the negative energy $C_{band}=+1$ band. More details for substrate effects on band structures can be found in Supplemental Material (SM).  Crucially,  deviations from the chiral limit  destroy band flatness, inducing the competing kinetic energy that influences correlated phases.

Let us now project onto the  $C_{band}=-2$ band, treating the charge neutrality point (occupied negative-energy bands) as the vacuum.  The projected Coulomb interaction~\cite{reddy2023fractional} reads
\begin{align}
H_{int}=\frac{1}{2A}\sum_{\boldsymbol\beta_{1,2,3,4}} & \sum_{\bold{k}_1\bold{k}_2\bold{q}}  V_{\boldsymbol\beta_1,\boldsymbol\beta_2,\boldsymbol\beta_3,\boldsymbol\beta_4}(\bold{k}_1, \bold{k}_2,\bold{q}) \times  \notag \\
&\psi_{\boldsymbol\beta_1,\bold{k}_1}^{\dagger}\psi_{\boldsymbol\beta_2,\bold{k}_2}^{\dagger}\psi_{\boldsymbol\beta_3,[\bold{k}_2-\bold{q}]}\psi_{\boldsymbol\beta_4,[\bold{k}_1+\bold{q}]},
\label{eq:Coulomb}
\end{align}
where $\psi_{\boldsymbol\beta,\bold{k}}^{\dagger}$ corresponds to the Bloch basis obtained from diagonalizing Eq.~\eqref{eq:H0}; flavor $\boldsymbol\beta_i=(\sigma_i,V_i)$ labels spin and valley degrees of freedom;  and $A=\frac{8\pi^2 N_s}{3\sqrt{3} k_{\theta}^2}$ is the system area with $N_s$ moir\'e cells.
The momentum sum is restricted to the moir\'e first Brillouin zone (FBZ), with $[\cdot]$ indicating momentum reduction to the FBZ via reciprocal lattice vector shifts. The interaction matrix element is related to the form factor via
$V_{\boldsymbol\beta_1,\boldsymbol\beta_2,\boldsymbol\beta_3,\boldsymbol\beta_4}(\bold{k}_1, \bold{k}_2,\bold{q})=V(\bold{q})\langle \boldsymbol\beta_1,\bold{k}_1|e^{-i\bold{q} \cdot \bold{r}}|\boldsymbol\beta_4,[\bold{k_1}+\bold{q}]\rangle \langle \boldsymbol\beta_2,\bold{k}_2|e^{i\bold{q} \cdot \bold{r}}|\boldsymbol\beta_3,[\bold{k_2}-\bold{q}]\rangle$. Here $V(\bold q)=\frac{2\pi e^{2}k_0}{|\bold q| \epsilon}(1-\delta_{\bold{q},0})$ is the Coulomb potential with $k_0$ being the Coulomb constant, and the dielectric constant is fixed to $\epsilon=4$. 

Coulomb interactions in twisted graphene can lift the four-fold flavor degeneracy of Chern bands~\cite{cao2018magic,zondiner2020cascade,wong2020cascade,xia2025topological,kwan2024strong,kwan2024fractional}. Focusing on the $C_{band}=-2$ band, we consider two distinct regimes of flavor polarizations as illustrated in Fig.~\ref{fig:band} (b). Case (i): For $0<\nu<1$, only one flavor $\boldsymbol\beta =(\sigma,K)$ is populated. The projected many-body Hamiltonian simplifies to 
\begin{align}
H_{\boldsymbol\beta}^{0<\nu<1}=P_{\boldsymbol\beta}(H_0+H_{int})P_{\boldsymbol\beta},
\label{eq:H_01}
\end{align}
where $P_{\boldsymbol\beta}$ projects onto the flavor $\boldsymbol\beta$. The first term corresponds to the dispersion of the non-interaction band, and the projected Coulomb interaction corresponds to intra-flavor scattering. Case (ii): At $\nu_{\rm{total}}=3+\nu$ with maximal inter-flavor interaction. we consider three fully filled flavors $(\bar{\sigma},K), (\sigma,K'), (\bar{\sigma},K')$ as a background. The interaction between the three filled flavors and the remaining  $\boldsymbol\beta=(\sigma,K)$ flavor is given by a mean-field Hamiltonian
\begin{align}
 H_{\boldsymbol\beta}^{MF}&=\frac{1}{2A} \sum_{\boldsymbol\beta'\neq \boldsymbol\beta} \sum_{\bold{k}\bold{k}'\bold{G}}  \hat{n}_{\boldsymbol\beta,\bold{k}} \langle\hat{n}_{\boldsymbol\beta',\bold{k}'}\rangle \notag \\ 
  & \times (V_{\boldsymbol\beta,\boldsymbol\beta',\boldsymbol\beta',\boldsymbol\beta}(\bold{k}, \bold{k}',\bold{G})+ V_{\boldsymbol\beta',\boldsymbol\beta,\boldsymbol\beta,\boldsymbol\beta'}(\bold{k}', \bold{k},\bold{G}) ),
\label{eq:H_mf}
\end{align}
where $\bold{G}$ is reciprocal lattice vector, $\hat{n}_{\boldsymbol\beta,\bold{k}}=\psi^{\dag}_{\boldsymbol\beta,\bold{k}}\psi_{\boldsymbol\beta,\bold{k}}$,  and $\langle \hat{n}_{\boldsymbol\beta',\bold{k}'} \rangle =1$ for the three background $\boldsymbol\beta'$ flavors.
Note that here only the Hartree term is present since form factors of the Fock term between different flavors vanish.  Then the full Hamiltonian reads
\begin{align}
H_{\boldsymbol\beta}^{\nu_{\rm{total}}=3+\nu}=P_{\boldsymbol\beta}(H_0+ H_{\boldsymbol\beta}^{MF} + H_{int} )P_{\boldsymbol\beta}.
\label{eq:H_34}
\end{align}
Here $H_0+ H_{\boldsymbol\beta}^{MF}$ represents the mean-field renormalized kinetic energy; see the red curve in Fig.~\ref{fig:band} (a). The single-particle band width is significantly reduced from $\sim 20$ meV to $\sim 15$ meV and may potentially stabilize correlated Hall states over a broader parameter range. For intermediate fillings $1<\nu_{\rm{total}}<3$, similar but weaker Hartree effects occur, so we focus on the two extremal regimes above.

To investigate correlated Hall states, we perform exact diagonalization which handles intra-flavor scattering in Eqs.~\eqref{eq:H_01} and~\eqref{eq:H_34} exactly.  The momentum grid is discretized as $\bold{k}=k_1 \mathbf{T}_1+k_2 \mathbf{T}_2$, where $\mathbf{T}_{1(2)}$ are unit momentum vectors, $k_{1(2)}=0,1,2,...,N_{1(2)}-1$ labels the coordinate, and $N_s=N_{1}N_{2}$ labels system size~\cite{SM}. For filling fraction $\nu=N_p/N_s$ with $N_p$ filled electron, the total momentum of occupied electrons $\mathbf{k}=\sum_{i=1}^{N_p}\mathbf{k}_i$ is a good quantum number, and the 2D coordinate $(k_1,k_2)$ is mapped to the quasi-1d index $k_1+N_1k_2$ for convenience. In the following, we analyze gapped Hall states focusing on flavor polarization $\nu_{\rm{total}}=3+\nu$.

\begin{figure}[hbt!]
\begin{localgraphicspath}{{figs/draft_fig/}}
\includegraphics[width=1\columnwidth]{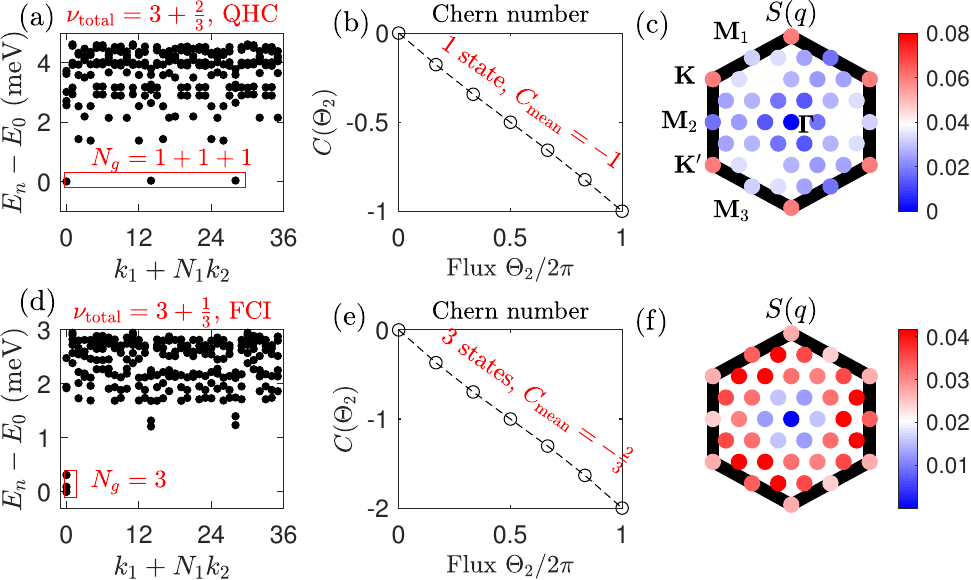}
\caption{Particle-hole asymmetry: \textbf{Quantum Hall crystal} at $\nu_{\rm{total}}=3+2/3$ in (a)-(c) and \textbf{fractional Chern insulator} at $\nu_{\rm{total}}=3+1/3$ in (d)-(f) on $N_s=36$ cluster with $w_{AA}/w_{AB}=0.6$. Three columns show energy spectrum, Chern number and structure factor $S(\mathbf{q})$.}
\label{fig:QHC_FCI}
\end{localgraphicspath}
\end{figure}

{\bf \emph{Quantum Hall crystal} at $\nu=\frac{2}{3}$.}~We first present evidence of the Hall crystal state
 at band filling $2/3$, as shown for an $N_s=36$ cluster in Fig.~\ref{fig:QHC_FCI} (a)-(c). We observe an interaction-induced many-body spectral gap in Fig.~\ref{fig:QHC_FCI} (a) with $N_g=1+1+1=3$ fold ground state degeneracy. The ground states reside in three momentum sectors $\boldsymbol{\Gamma},\mathbf{K},\mathbf{K}'$  independent of the cluster geometry.  These ground state momenta are different from the FCIs~\cite{regnault2011fractional}.
To understand the topological property of this state, we compute the many-body Chern number \cite{niu1985quantized,sheng2003,fukui2005Chern} which connects to the Hall transport through $\sigma_{H}=\frac{e^2}{h}C_{mean}$ with $C_{mean}=\sum_{i=1}^{N_g} C_i/N_g$. Here the Chern number for a single many-body ground state $|\Phi_i\rangle$ is defined as an integral over twisted boundary conditions $(\theta_1,\theta_2)$:
\begin{align}
C_{i}&=C_{i}(2\pi)-C_{i}(0), \notag\\
C_{i}(\Theta_2)&=\frac{1}{2\pi}\int_{0}^{\Theta_2}d\theta_2 \int_{0}^{2\pi}d\theta_1 F_{i}(\theta_1,\theta_2),
\label{eq:Chern}
\end{align}
where  the many-body berry curvature takes the form $F_{i}(\theta_1,\theta_2)=i(\langle \partial_{\theta_1}\Phi_{i}|\partial_{\theta_2}\Phi_{i} \rangle-\langle \partial_{\theta_2}\Phi_{i}|\partial_{\theta_1}\Phi_{i} \rangle)$. The Chern number $C_{mean}=-1$ for a representative momentum sector can be read from Fig.~\ref{fig:QHC_FCI} (b), where the minus sign is inherited from the band Chern number.
Unlike FCIs, the violation of  $C_{mean}=\nu C_{band}$ here further signals a non-FCI phase.
We then compute the static density-density structure factor
\begin{align}
S(\mathbf{q})=\frac{1}{N_s}[\langle \tilde{\rho}_{\mathbf{q}} \tilde{\rho}_{-\mathbf{q}}\rangle - \delta_{q,0}\langle \tilde{\rho}_{\mathbf{q}} \rangle \langle \tilde{\rho}_{-\mathbf{q}}\rangle],
\label{eq:Sq}
\end{align}
where $\tilde{\rho}_{\mathbf{q}} \approx   \sum_{k\in \text{FBZ}} \langle \mathbf{k}|e^{-i\mathbf{q}\cdot \mathbf{r}}|\mathbf{k+q}\rangle  \psi_{\mathbf{k}}^\dagger \psi_{\mathbf{k}+\mathbf{q}}$ is the projected density operator. The $S(\mathbf{q})$ in Fig.~\ref{fig:QHC_FCI} (c) exhibits sharp peaks at the $K,K'$ points, signaling a CDW order with $\sqrt{3}\times \sqrt{3}$ enlarged unit-cell. In the SM, we further corroborate this structural reorganization through the real-space pair correlation function. Together, the spectrum degeneracy, integer Chern number, and CDW order conclusively establish a $\nu=2/3$ QHC phase~\cite{polshyn2022topological,pan2022topological,xu2024maximally, song2024intertwined,  sheng2024quantum,dong2024anomalous,waters2024interplay,lu2025extended}. 
We also note that a similar crystal phase was recently predicted in twisted double bilayer graphene which also hosts high Chern bands~\cite{perea2024quantum}. 
It is interesting to investigate the possible presence of higher Chern bands in other moir\'e systems where $\nu=2/3$ QHCs have been observed~\cite{waters2024interplay,su2025moire}.

{\bf \emph{Fractional Chern insulator at $\nu_{\rm{total}}=3+\frac{1}{3}$.}}~Remarkably, a translation-invariant FCI emerges at $\nu=1/3$, indicating dramatic particle-hole asymmetry for correlated states in this higher Chern band system.  The energy spectrum in Fig.~\ref{fig:QHC_FCI} (d) shows three-fold gapped ground states with momentum counting different from the QHC in Fig.~\ref{fig:QHC_FCI} (a). The total Chern number $\sum{C_i}=-2$ for the three degenerate ground states [Fig.~\ref{fig:QHC_FCI} (e)] yields a fractional Hall conductance $|\sigma_{H}|=2e^2/3h$. Moreover, the absence of sharp structure factor peaks [Fig.~\ref{fig:QHC_FCI} (f)] demonstrates that the real space moir\'e translation symmetry remains unbroken. 
To further characterize the topological order, we analyze the particle entanglement spectrum (PES), which encodes quasihole statistics~\cite{li2008entanglement,regnault2011fractional,sterdyniak2011extracting}.
The PES fingerprint (see SM) matches the Halperin $(112)$ state~\cite{halperin1983theory,geraedts2015competing,dong2023many,liu2024engineering}, a two-component pseudospin singlet quantum Hall state.
This identification aligns with the Chern number decomposition $C_{band}=(-1)+(-1)$, analogous to spinful Landau levels.
The emergence of multicomponent FCI state in the higher Chern band opens pathways to engineer non-Abelian anyons via coupling to superconductors~\cite{mong2014universal}, offering a tantalizing platform for topological quantum computation.

\begin{figure}[hbt!]
\begin{localgraphicspath}{{figs/draft_fig/}}
\includegraphics[width=0.9\columnwidth]{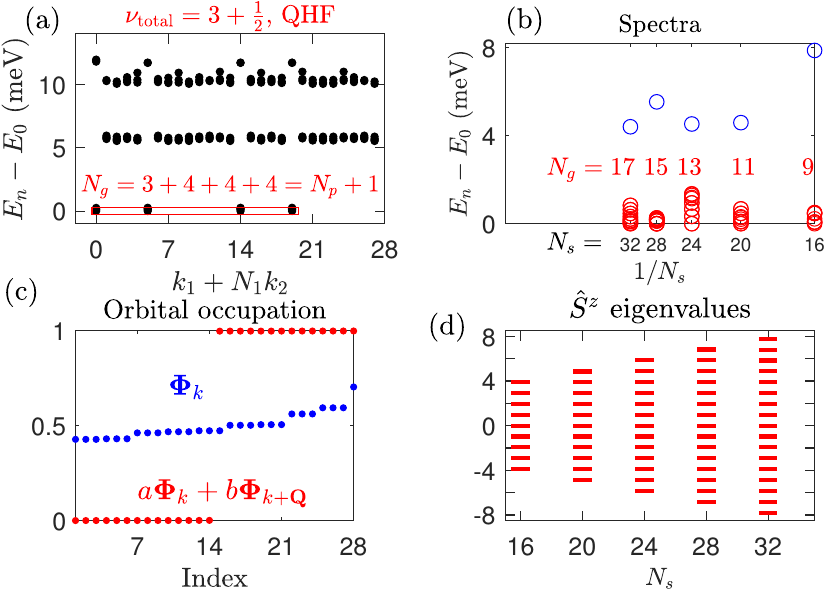}
\caption{{\bf Quantum Hall ferromagnet at $\nu_{\rm{total}}=3+1/2$ with $w_{AA}/w_{AB}=0.6$.}  (a) Many-body energy spectrum on the $N_s=28$ cluster.  (b) Degeneracy and spectrum gap across different cluster sizes. (c) Typical orbital occupation numbers of the momentum eigenstate (blue) and $\mathbf{Q}$-superposed CDW state (red) on the $N_s=28$ cluster. (d) Emergent $S^z$ quantum numbers for the pseudospin ferromagnetism with $S^{z}\approx -N_p/2,-N_p/2+1,...,N_p/2-1,N_p/2$. }
\label{fig:QHF}
\end{localgraphicspath}
\end{figure}

{\bf \emph{Quantum Hall ferromagnet at $\nu_=\frac{1}{2}$.}}~At half filling, we uncover a distinct gapped phase characterized by an extensive ground state degeneracy, contrasting sharply with the fixed degeneracies of QHCs or FCIs. For instance, on the $N_s=28$ cluster, we observe $N_g=3+4+4+4$ quasi-degenerate states in the four momentum sectors $\boldsymbol{\Gamma},\mathbf{M}_1=\mathbf{G}_2/2,\mathbf{M}_2=\mathbf{G}_1/2$ and $\mathbf{M}_3=(\mathbf{G}_1+\mathbf{G}_2)/2 $ [Fig.~\ref{fig:QHF} (a)]. 
As shown in Fig.\ref{fig:QHF}(b), different clusters exhibit a degeneracy $N_g=N_p+1$ scaling with system size, with surprisingly small energy splitting ($\Delta_{E_g}<0.5 {\rm meV}$ on symmetric clusters) compared to the $\sim 15 {\rm meV}$ bandwidth. 
This unusual degeneracy points towards a QHF state~\cite{ezawa2009quantum,dong2023many,wilhelm2023non}, of which the extensive degeneracy arises from an emergent pseudospin $\rm{SU}(2)$ symmetry.  

Reference~\onlinecite{dong2023many} shows that an ideal $|C_{band}|=2$ flat band can be mapped to a pair of spinful $|C_{band}|=1$ bands. In this framework, pseudospin ferromagnetism predicts $N_g=2S+1=N_p+1$ fold degeneracy, quantized many-body Chern number $|C_{mean}|=1$ and CDW orders.  We confirm that the average Chern number is indeed $-1$, with the structure factor showing sharp peaks at $\mathbf{M}_{1,2,3}$ points indicating CDW order. However, the above evidence does not prove the existence of the $\rm{SU}(2)$ structure directly for our system, since the Chern band mapping method is devised for the ideal flat band and does not include effects of Coulomb interactions.

To justify the existence of emergent $\rm{SU}(2)$ symmetry, here we develop a generic method to unravel the pseudospin ferromagnetism in the presence of band dispersion and interaction; for details, see SM. Motivated by  the distribution of ground state momenta $\mathbf{k}\in \{\mathbf{\Gamma},\mathbf{M}_1,\mathbf{M}_2,\mathbf{M}_3\}$, we construct CDW states with ordering wavevector $\mathbf{Q} \in \{ \mathbf{M}_1, \mathbf{M}_2, \mathbf{M}_3 \}$ via linear superposition of momentum eigenstates:
\begin{align}
\tilde{\boldsymbol{\Phi}}_{\mathbf{k},\mathbf{k}+\mathbf{Q}}= \sum_{i}a_{i} |\boldsymbol{\Phi}^{i}_{\mathbf{k}}\rangle +\sum_{j}b_{j} |\boldsymbol{\Phi}^{j}_{\mathbf{k}+\mathbf{Q}}\rangle,
\label{eq:psi_two_component}
\end{align}
where coefficients $\{a_i,b_j\}$ are to be determined. The CDW order parameter can be extracted from expectation values $\langle\psi_{\mathbf{k}}^{\dagger}\psi_{\mathbf{k}+\mathbf{Q}}\rangle$. Crucially, such superposition approximately forms psueospin polarized state, indicated by eigenvalues of the momentum space correlation matrix
\begin{align}
O_{m,n}=\langle \psi_{\mathbf{k}_m}^{\dagger}\psi_{\mathbf{k}_n}\rangle.
\end{align}
For momentum eigenstates $\boldsymbol{\Phi}^{i}_{\mathbf{k}}$, the $O$ matrix is diagonal, and its eigenvalues simply return the momentum distribution $n_k$ [see blue dots in Fig.~\ref{fig:QHF} (c)]. The $n_k$'s are almost independent of the ground state labels  $\mathbf{k},i$, which explains the suppressed (kinetic) energy splitting. On the other hand, for superposed $\tilde{\boldsymbol{\Phi}}_{\mathbf{k},\mathbf{k}+\mathbf{Q}}$ states the $O$ matrix reduces to $2\times 2$ blocks connected by $\mathbf{Q}$. Focusing on $\mathbf{Q}=\mathbf{M}_2$, through optimizing $\{a_i,b_j\}$, we obtain a product-state-like wavefunction of which the orbital occupations (eigenvalues) are almost $0$ or $1$ [see red dots in Fig.~\ref{fig:QHF} (c)].  Thus we have demonstrated $\tilde{\boldsymbol{\Phi}}_{\mathbf{k},\mathbf{k}+\mathbf{Q}}\approx \prod_{k\in\text{rFBZ}} \tilde{\boldsymbol{\phi}}_{0,\mathbf{k}}^{\dagger}$ is almost a Slater-determinant~\cite{note_slater_determinant} with $\tilde{\boldsymbol{\phi}}_{0,\mathbf{k}}^{\dagger}=\alpha_{k}\psi_k^{\dagger}+\beta_{k}\psi_{k+\mathbf{G}_1/2}^{\dagger}$ being the fully occupied orbitals obtained from diagonalizing $O$. 
 
We are now able to define pseudospin operators $\hat{S}^{x/y/z} =\sum_{\mathbf{k}\in \text{rFBZ}} \hat{s}^{x/y/z}_{\mathbf{k}}$ in the reduced Brillouin zone using $\tilde{\boldsymbol{\phi}}_{0,\mathbf{k}}^{\dagger}$ as the south pole and another symmetry related orbital set $\tilde{\phi}_{1,\mathbf{k}}^{\dagger}  =\alpha_{\mathbf{k}+\frac{\mathbf{G}_{2}}{2}}\psi_{\mathbf{k}+\frac{\mathbf{G}_{2}}{2}}^{\dagger}- \beta_{\mathbf{k}+\frac{\mathbf{G}_{2}}{2}}\psi_{\mathbf{k}+\frac{\mathbf{G}_{1}}{2}+\frac{\mathbf{G}_{2}}{2}}^{\dagger}$ as the north pole:
\begin{align}
\hat{s}^{x/y/z}_{\mathbf{k}}=\tilde{\boldsymbol{\phi}}_{\mathbf{k}}^{\dagger}\sigma^{x/y/z}\tilde{\boldsymbol{\phi}}_{\mathbf{k}}/2,\,\,\,\, \tilde{\boldsymbol{\phi}}_{\mathbf{k}}^{\dagger}=[\tilde{\phi}_{0,\mathbf{k}}^{\dagger},\tilde{\phi}_{1,\mathbf{k}}^{\dagger}].
\end{align}
An $\rm{SU}(2)$ rotation on the south pole realizes a general ferromagnetic state constructed as $\boldmath{\Phi}(\varphi,\phi)=\prod_{\mathbf{k}}(\cos\frac{\varphi}{2} \tilde{\phi}_{0,\mathbf{k}}^{\dagger}+\sin\frac{\varphi}{2}e^{i\phi} \tilde{\phi}_{1,\mathbf{k}}^{\dagger})$. One can check that the CDW wavevector $\mathbf{Q}$ is locked to the polarization on the Bloch sphere. We remark that the emergent orbitals $\{\alpha_{k},\beta_{k}\}$ are determined from a full realistic Hamiltonian instead of the flat-band model in Ref.~\onlinecite{dong2023many}.
The spectrum of the $N_g\times N_g$ operator $\hat{S}^z$ is shown in Fig.~\ref{fig:QHF} (d), where the nearly quantized eigenvalues (with quantization error $\delta S^z<3\%$) conclusively justify that the $N_g$-fold ground states form a spin $S=N_p/2$ multiplet. Notably, the error of $S^z$ remains unchanged across a broad parameter range $w_{AA}/w_{AB}<0.8$~\cite{SM}, suggesting that the emergent $\rm{SU}(2)$ symmetry is not fine-tuned and always holds approximately before the transition in our model. The strongly suppressed ground state energy splitting and the nearly quantized $S^z$ quantum numbers demonstrate that the QHF state is quite robust and exists beyond the chiral limit.


\begin{figure}[hbt!]
\begin{localgraphicspath}{{figs/draft_fig/}}
\includegraphics[width=1\columnwidth]{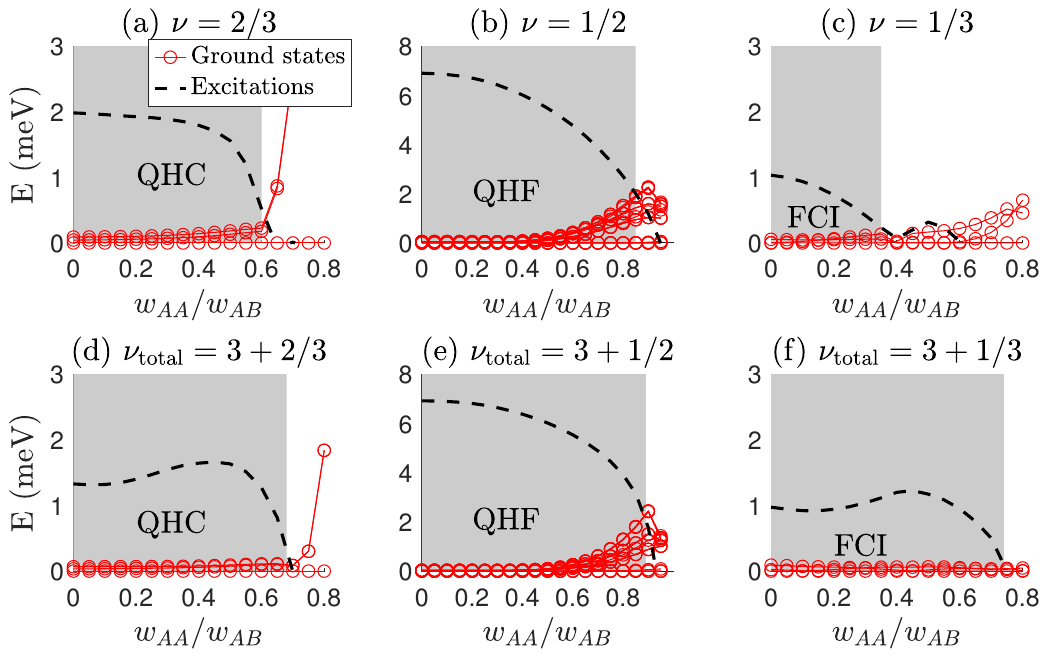}
\caption{{\bf Energy spectra and quantitative phase diagrams versus $w_{AA}/w_{AB}$.} Shaded regions show ranges of gapped ground states. Red curves show the evolution of $N_g$ quasi-degenerate ground states, while the black curves show the lowest excitation energies.  System sizes for simulating the three phases are $N_s=27,28,24$, on which the degeneracies are $N_g=3,15,3$, respectively.}
\label{fig:phase_diagram}
\end{localgraphicspath}
\end{figure}

{\bf \emph{Global phase diagram.}}~Having identified the three gapped anomalous Hall states, we examine their stability across the full range of lattice corrugation parametrized through  $w_{AA}/w_{AB}$. The essential control parameter is the interaction-to-bandwidth ratio $U/W$, since Coulomb-driven gaps can only open if $U/W$ is sufficient large.
Panels (a-c) of Fig.~\ref{fig:phase_diagram} show that, for the low filling regime $\nu\in(0,1)$, all three phases are stabilized close to the chiral limit $w_{AA}=0$, where the single-particle bandwidth is minimal and $U/W$ is maximal.
The critical threshold for the FCI state at $\nu=1/3$ is about $w_{AA}\approx 0.35 w_{AB}$, much lower than the expected realistic value. 
By contrast, in the high-filling regime $\nu_{\rm{total}}=3+\nu$, interband interaction renormalizes the dispersion and thereby boosts $U/W$. As panels (d-f) demonstrate, this enhanced interaction strength stabilizes all three phases over a much wider corrugation window. In particular, the parameter range of the FCI at $\nu_{\rm{total}}=3+1/3$ is significantly extended compared to its $\nu=1/3$ counterpart.
The magnitudes of gaps in the overall phase diagrams indicate that the FCI state is relatively more fragile than the other two states, suggesting that insufficiently low temperature may account for the absence of experimental signatures at $\nu=1/3$. 
In contrast, the QHF phase exhibits the largest gap and the widest stability window, suggesting that it is the most accessible for observation.

{\bf \emph{Summary.}}~Our work predicts a series of novel gapped Hall states at fractional filling of HTG subjected to a substrate potential. The phase diagram reveals three distinct phases in realistic higher Chern bands beyond the chiral limit: a quantum Hall crystal, a Halperin fractional Chern insulator, and an emergent $\rm{SU}(2)$  Hall ferromagnet. Each phase exhibits either unique symmetry-breaking orders or topological orders compared to those expected in the $|C_{band}|=1$ band,  and they can be differentiated by their Hall conductance or CDW ordering pattern in  real- and momentum-space. Our study also makes significant progress in understanding the emergent $\rm{SU}(2)$ symmetry. The construction of pseudospin operators and quantitative computation of quantum numbers open the way to resolving emergent symmetries in generic interacting, dispersive Chern bands.
The filling fractions $\nu_{\rm{total}}=2/3,7/2$ highlighted in the recent transport experiment \cite{xia2025topological} partially align with our predicted doping sequence. 
Future low-temperature transport measurements at the optimal carrier densities  $\nu_{\rm{total}}=3+\nu$ may further
clarify the nature of the quantum states observed in experiments.

{\bf \emph{Acknowledgment}}.~This work is supported by the US National Science Foundation (NSF) Grant No. PHY-2216774.
Additional support was provided by the Caltech Institute for Quantum Information and Matter, an NSF
Physics Frontiers Center (NSF Grant PHY-2317110). The numerical simulation is supported by NSF instrument grant  DMR-2406524. 

\bibliography{draft.bib}

\begin{thebibliography}{62}
\bibitem [{not()}]{note_slater_determinant}%
  \BibitemOpen
  \href@noop {} {\bibinfo  {journal} {For any many-body state, the eigenvalue
  of correlation matrix is always bounded between $0$ and $1$. If the state is
  a correlated non-Slater-determinant state, the eigenvalues of correlation
  matrix are away from $0$ and $1$. On the other hand, if the eigenvalues reach
  $0$ or $1$, then the wavefunction is a Slater-determinant with the
  single-particle orbitals being eigenvectors of the correlation matrix}\
  }\BibitemShut {NoStop}%
\end{thebibliography}

\end{document}


\newcommand{\dagga}{{\phantom{\dagger}}}
\newcommand{\SN}[1]{{\color{red}[SN: #1]}}
\newcommand{\JA}[1]{\textcolor{blue}{(JA) #1}}

\renewcommand{\theequation}{S\arabic{equation}}
\renewcommand{\thefigure}{S\arabic{figure}}
\renewcommand{\thetable}{S\Roman{table}} 

\renewcommand*{\citenumfont}[1]{S#1}
\renewcommand*{\bibnumfmt}[1]{[S#1]}

\title{Supplemental materials}
\maketitle
\onecolumngrid

\section{Effects of $w_{AA}$ and substrate potential on band structures}
\begin{figure}[hbt!]
\begin{localgraphicspath}{{figs/draft_fig/}}
\includegraphics[width=0.6\columnwidth]{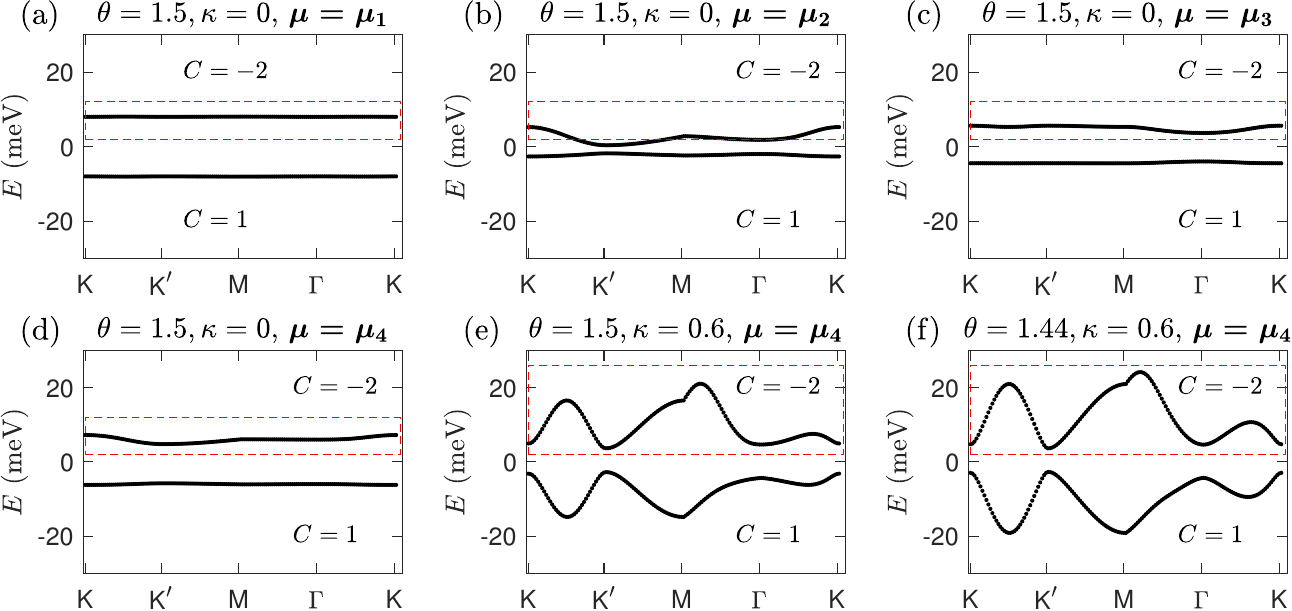}
\caption{Non-interacting Chern bands for different $\theta, \kappa=w_{AA}/w_{AB}$ and substrate potential $\boldsymbol{\mu}$. Here $\boldsymbol{\mu}_1=(-8,-8,-8)$ meV corresponds to a uniform potential, $\boldsymbol{\mu}_2=(0,0,-20)$ meV is a single-layer potential, $\boldsymbol{\mu}_3=(-8,0,-8)$ meV is a potential acting on both bottom and tom layers, and $\boldsymbol{\mu}_4=(-4,-6,-8)$ meV is a potential that decays from bottom layer to top layer which is used in the main text.}
\label{fig:band}
\end{localgraphicspath}
\end{figure}

It is known that in the chiral limit \cite{guerci2024chern} and at the magic angle $\theta=1.5^{\circ}$, the two central Chern bands with $C_{band}=+1,-2$ have exact zero energies at the $K$ valley. As the two bands are eigenstates of the sublattice basis,  a uniform substrate potential $\boldsymbol{\mu}_1=(-8,-8,-8)$ meV can separate the two bands perfectly without changing the dispersion; see Fig.~\ref{fig:band} (a). Due to the sign of the potential, the $C_{band}=-2$ band has positive energy and the $C_{band}=+1$ band has negative energy. However, such uniform potential is not realistic in real materials. Figure~\ref{fig:band} (b)-(d) explores different types of non-uniform substrate potentials including a single-layer potential $\boldsymbol{\mu}_2=(0,0,-20)$ meV, a potential acting on both top and bottom layers $\boldsymbol{\mu}_3=(-8,0,-8)$ meV, and a potential decaying from the bottom layer to the top layer  $\boldsymbol{\mu}_4=(-4,-6,-8)$ meV. One can see that the two bands separate most effectively for $\boldsymbol{\mu}_3$ and $\boldsymbol{\mu}_4$; we stick to $\boldsymbol{\mu}_4$ throughout this work but expect $\boldsymbol{\mu}_3$ (and possibly also $\boldsymbol{\mu}_2$) to exhibit similar phenomenology.  Figure~\ref{fig:band} (e) further considers realistic corrugation $w_{AA}=0.6w_{AB}$ which is far away from the chiral limit, yielding a more dispersive band with a bandwidth close to $20$ meV. In Fig.~\ref{fig:band} (f), one can see that at a slightly different twist angle $\theta=1.44^{\circ}$ used in main text,  the band structures are similar. In the opposite valley $K'$, the signs of Chern numbers reverse, i.e., the upper band has $C_{band}=+2$ and the lower band has $C_{band}=-1$.

\section{Finite clusters for exact diagonalization}
We denote $\mathbf{a}_1,\mathbf{a}_2$ as the lattice vectors of the moiré trianglular superlattice. The reciprocal lattice vectors takes the form
\begin{align}
\mathbf{g}_i=\frac{2\pi\epsilon_{ij}\mathbf{a}_j\times\hat{z}}{|\mathbf{a}_1\times\mathbf{a}_2|}.
\end{align}
A finite cluster can be determined from real space translation vectors $\mathbf{L}_1=m_1 \mathbf{a}_1+n_1 \mathbf{a}_2$, $\mathbf{L}_2=m_2 \mathbf{a}_1+n_2 \mathbf{a}_2$, where vectors $\mathbf{L}_1,\mathbf{L}_2$ define the translation invariance of the finite cluster on an infinite lattice. The allowed plane-wave momenta take the form
\begin{align}
\mathbf{T}_i=\frac{2\pi\epsilon_{ij}\mathbf{L}_j\times\hat{z}}{|\mathbf{L}_1\times\mathbf{L}_2|},
\end{align}
which satisfies $\mathbf{L}_i \cdot \mathbf{T}_j =2\pi \delta_{ij}$. The discrete momentum points take the form of integer combinations $\mathbf{k}=k_1 \mathbf{T}_1+k_2\mathbf{T}_2$, where $k_i=0,1,...,N_i-1$ and $N_1 N_2=N_s$. In our many-body calculations, we use the total momentum of occupied electrons $\mathbf{k}=\sum_{i=1}^{\nu N_s}\mathbf{k}_i$ as a symmetry to block diagonalize the many-body Hamiltonian. For each cluster the total number of $\mathbf{k}$ sectors equals to the cluster size $N_s$. 

In Table \ref{table:clusters}, we show information for all the clusters used in numerical calculations. Note that since the quantum Hall crystals at $\nu=2/3$ have ground state momenta $\Gamma,K,K'$, numerical diagonalization on clusters without $\mathbf{K},\mathbf{K}'$ is biased. Therefore, among the two $N_s=24$ clusters in the table, we take the the one with  $\mathbf{L}_1=(1,4)$, $\mathbf{L}_2=(5,-4)$ for calculations at $\nu=2/3$, and take the one with $\mathbf{L}_1=(4,-4)$, $\mathbf{L}_2=(0,6)$ otherwise.

\begin{table}
\begin{tabular}{|c|c|c|c|c|c|c|}
\hline
  $N_s$ & $\mathbf{L}_1$ & $\mathbf{L}_2$ & $N_1$ & $N_2$ & $\mathbf{K},\mathbf{K}'$ points \tabularnewline
\hline
  16 & $(4,0)$ & $(0,4)$ &4&4& \text{No}  \tabularnewline
\hline
  20 & $(2,2)$ & $(4,-6)$ &2&10& \text{No}  \tabularnewline
\hline
  24 & $(4,-4)$ & $(0,6)$ &4&6& \text{No}  \tabularnewline
\hline
  24 & $(1,4)$ & $(5,-4)$ &1&24& \text{Yes} \tabularnewline
\hline
  26 & $(0,13)$ & $(2,4)$ &13&2& \text{No} \tabularnewline
\hline
27 & $(6,-3)$ & $(3,-6)$ &3&9& \text{Yes} \tabularnewline
\hline
28 & $(-2,6)$ & $(4,2)$ &2&14& \text{No} \tabularnewline
\hline
30   & $(5,0)$&$(0,6)$ &5&6& \text{No}\tabularnewline
\hline
32 & $(2,4)$ & $(6,-4)$ &2&16& \text{No} \tabularnewline
\hline
36 & $(6,0)$ & $(0,6)$ &6&6& \text{Yes} \tabularnewline
\hline
\end{tabular}\caption{\label{table:clusters} Information about finite clusters used in ED.}
\end{table}

\section{Further numerical analysis for quantum Hall ferromagnets }
In the main text, based on the observations of the extensive ground state degeneracy and the presence of $\mathbf{M}$ point structure factor peaks, we propose that the half-filled ground states are quantum Hall ferromagnets. To further substantiate this proposition, we analyze the symmetry-breaking charge density wave (CDW) orders and identify signatures of emergent $\rm{SU}(2)$ symmetry, including pseudospin ferromagnetism and suppressed ground state energy splitting.

\subsection{Symmetry breaking orders in quantum Hall ferromagnets}
As demonstrated in Ref.~\onlinecite{dong2023many}, ideal Chern bands permit the construction of many-body ground states resembling mean-field product states. For our non-ideal band under realistic Coulomb interactions, we quantify the deviation of a generic many-body state $\tilde{\mathbf{\Phi}}$ from a product state using the momentum-space correlation matrix:
\begin{align}
O_{i,j}=\langle \tilde{\mathbf{\Phi}}|\psi_{\mathbf{k}_i}^{\dagger}\psi_{\mathbf{k}_j}|\tilde{\mathbf{\Phi} }\rangle,
\end{align}
where $\psi_{\mathbf{k}_i}^{\dagger}$ creates a Bloch state at momentum $\mathbf{k}_i$. In product states, $O$ has eigenvalues strictly 0 (unoccupied) or 1 (occupied), while interacting states exhibit eigenvalues distributed continuously between 0 and 1. 

\begin{figure*}[hbt!]
\begin{localgraphicspath}{{figs/draft_fig/}}
\includegraphics[width=0.7\columnwidth]{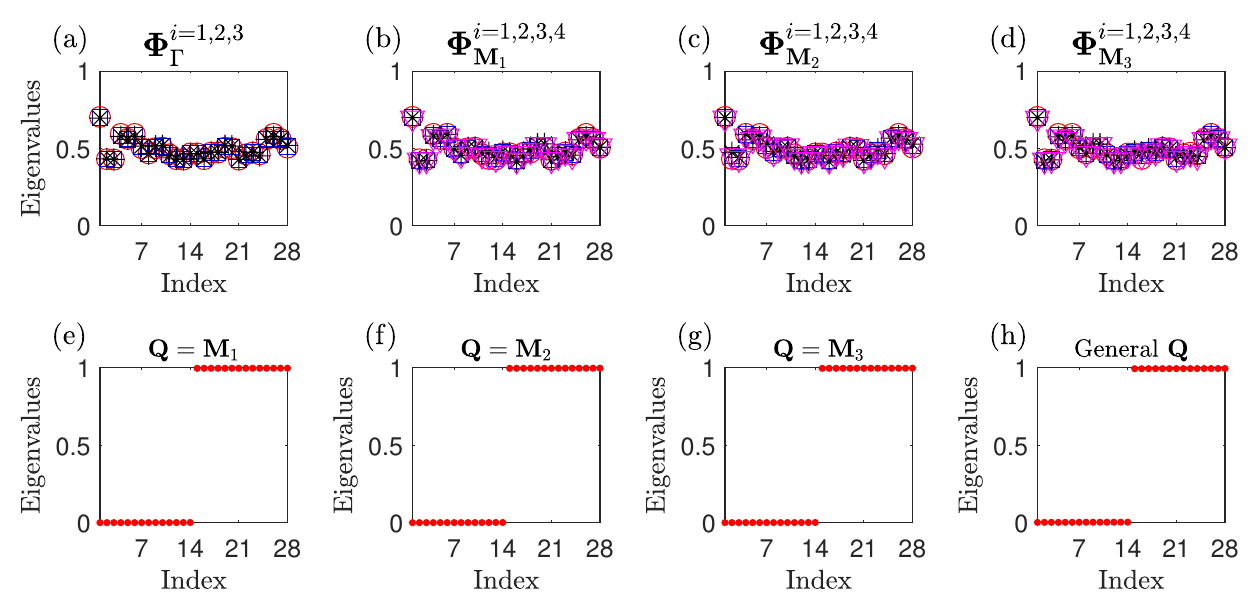}
\caption{Orbital occupation number revealed by eigenvalues of momentum space correlation matrix $\langle \psi_{k_1}^{\dagger}\psi_{k_2} \rangle$ with $\nu_{\rm{total}}=3+1/2, w_{AA}=0.6 w_{AB}, N_s=28$. (a)-(d) show the results for the $3,4,4,4$-fold ground states on the four momentum sectors $\mathbf{\Gamma},\mathbf{M}_1, \mathbf{M}_2, \mathbf{M}_3$, resectively.  Since the exact eigenstates are momentum conserving, the correlation matrices are diagonal and the elements simply correspond to Bloch momentum distribution $\langle  \psi_{\mathbf{k}}^{\dagger}\psi_{\mathbf{k}}\rangle$.   (e)-(g) show the results for mean-field like states obtained from linear combinations of states with different momenta using Eq.~\eqref{eq:psi_two_component}. (h) shows a general linearly combined state using Eq.~\eqref{eq:psi_general}.  For (a)-(d) the indices $1,...,N_s$ label momentum, whereas in (e)-(h), the indices $1,...,N_s$ refer to the order of eigenvalues sorted in ascending order. }
\label{fig:correlation_matrix}
\end{localgraphicspath}
\end{figure*}

Figures~\ref{fig:correlation_matrix} (a)--(d) display the eigenvalues of $O$ for the $N_g$-fold degenerate momentum-resolved ground states $|\mathbf{\Phi}_{\mathbf{k}}^i\rangle$, with total momentum $\mathbf{k} \in \{\mathbf{\Gamma}, \mathbf{M}_1, \mathbf{M}_2, \mathbf{M}_3\}$ and $i$ labeling states within each momentum sector. Momentum conservation restricts $O$ to be diagonal, where the diagonal elements $\langle \hat{n}_{\mathbf{k}} \rangle \equiv \langle \psi_{\mathbf{k}}^{\dagger} \psi_{\mathbf{k}} \rangle$ represent the momentum distribution. The observed deviation of $\langle n_{\mathbf{k}} \rangle$ from $0$ and $1$ confirms that these states are strongly correlated and distinct from product states.

Inspired by mean-field analysis in Ref.~\onlinecite{wilhelm2023non}, a general symmetry-broken state can be represented by linear superposition from distinct momentum sectors:
\begin{align}
\tilde{\boldsymbol{\Phi}}= \sum_{i}a_{i} |\boldsymbol{\Phi}^{i}_{\boldsymbol{\Gamma}}\rangle +\sum_{j}b_{j} |\boldsymbol{\Phi}^{j}_{\mathbf{M}_1}\rangle  +\sum_{k}c_{k} |\boldsymbol{\Phi}^{k}_{\mathbf{M}_2}\rangle +\sum_{l}d_{l} |\boldsymbol{\Phi}^{l}_{\mathbf{M}_3}\rangle,
\label{eq:psi_general}
\end{align}
where $a_i,b_j,c_k,d_l$ are parameters to be determined. In the simplest scenario involving only two momentum sectors, we construct states with ordering wavevector $\mathbf{Q} \in \{ \mathbf{M}_1, \mathbf{M}_2, \mathbf{M}_3 \}$:
\begin{align}
\tilde{\boldsymbol{\Phi}}_{\mathbf{k},\mathbf{k}+\mathbf{Q}}= \sum_{i}a_{i} |\boldsymbol{\Phi}^{i}_{\mathbf{k}}\rangle +\sum_{j}b_{j} |\boldsymbol{\Phi}^{j}_{\mathbf{k}+\mathbf{Q}}\rangle.
\label{eq:psi_two_component}
\end{align}
This superposition generates off-diagonal correlations between $\mathbf{k}_i$ and $\mathbf{k}_i + \mathbf{Q}$, reducing the correlation matrix $O$ to block-diagonal $2\times 2$ form. 

As shown in Figs.~\ref{fig:correlation_matrix} (e)--(g), optimized coefficients $a_i, b_j$ in Eq.~\eqref{eq:psi_two_component} yield correlation matrix eigenvalues approaching 0 or 1 for all choices of $\mathbf{Q}$, confirming proximity to mean-field product states. The eigenvectors of $O$ define the (almost) fully occupied orbitals:
\begin{align}
\phi_{\mathbf{k}}^{\dagger}  = \alpha_{\mathbf{k}} \psi_{\mathbf{k}}^{\dagger} + \beta_{\mathbf{k}} \psi_{\mathbf{k} + \mathbf{Q}}^{\dagger},
\end{align}
which will be utilized in subsequent analysis. For the general superposition in Eq.~\eqref{eq:psi_general}, $O$ reduces to $4\times 4$ blocks while still admitting mean-field-like solutions [Fig.~\ref{fig:correlation_matrix} (h)]. We now characterize the symmetry-breaking orders of these linearly combined states.

\begin{figure*}[hbt!]
\begin{localgraphicspath}{{figs/draft_fig/}}
\includegraphics[width=0.5\columnwidth]{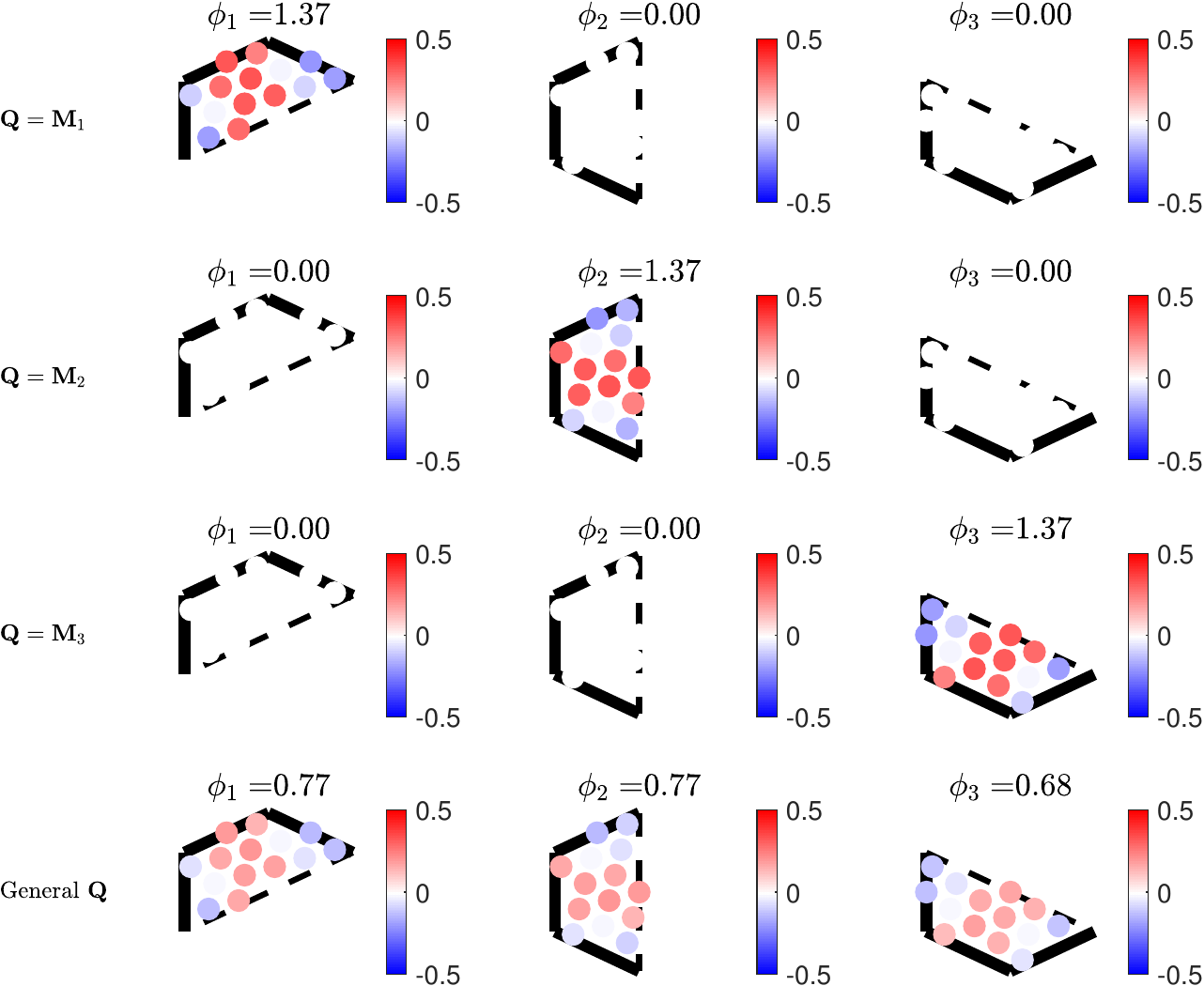}
\caption{Momentum space contributions to the charge density wave order parameter $\phi_{i}$ using Eq.~\eqref{eq:order}. Four different rows correspond to four different linearly combined mean-field-like states in Fig.~\ref{fig:correlation_matrix}(e)–(g). Three columns $\phi_1,\phi_2,\phi_3$ correspond to three different ways of folding Brillouin zone on the triangular lattice.  }
\label{fig:spin_texture}
\end{localgraphicspath}
\end{figure*}

To characterize CDW orders in the superposed many-body states, we compute order parameters following Refs.~\onlinecite{dong2023many,wilhelm2023non}:
\begin{align}
\phi_{i}= \frac{1}{2}\sum_{\mathbf{k} \in \text{folded BZ}}  \sum_{\xi=\pm 1}  F(\mathbf{k},\xi\mathbf{M}_i)\langle \tilde{\boldsymbol{\Phi}} |\psi_{\mathbf{k}}^{\dagger} \psi_{\mathbf{k}+\xi\mathbf{M}_i}| \tilde{\boldsymbol{\Phi}} \rangle +h.c.,
\label{eq:order}
\end{align}
where $F(\mathbf{k},\xi\mathbf{q})=\langle \mathbf{k}| e^{-i\mathbf{q}}|\mathbf{k+q}  \rangle$ is the form factor between Bloch states and $\mathbf{k}$  corresponds to the Bloch momentum in the folded Brillouin zone. 
Figure~\ref{fig:spin_texture} displays $\phi_i$ and their folded Brillouin Zone distributions for the optimized states in Figs.~\ref{fig:correlation_matrix} (e)-(h). When $\mathbf{Q} = \mathbf{M}_j$ (single-$Q$ states), only $\phi_j$ is nonzero by construction. These correspond to pseudospin-polarized states along the $x,y,z$ axes of an emergent Bloch sphere. 
For any solution $\{a_{i},b_{j}\}$ in Eq.~\eqref{eq:order}, the sign-flipped coefficients $\{a_i, -b_j\}$ yield an order parameter with opposite sign. 
The resulting state pairs $|\pm Z\rangle$ (similarly $|\pm X\rangle$, $|\pm Y\rangle$) represent opposite pseudospin polarizations along the same axis. For example, if $\mathbf{Q}=\mathbf{M}_2$, these have order parameters $(0, \pm\phi_2, 0)$. Their correlation matrices share identical diagonal elements but opposite off-diagonal signs. Consequently, if $|+Z\rangle$ has (almost) fully occupied orbitals
\begin{align}
\phi_{\mathbf{k}}^{\dagger} = \alpha_{\mathbf{k}} \psi_{\mathbf{k}}^{\dagger} + \beta_{\mathbf{k}} \psi_{\mathbf{k}+\mathbf{Q}}^{\dagger},
\end{align}
then $|-Z\rangle$ has 
\begin{align}
\phi_{\mathbf{k}}^{\dagger} = \alpha_{\mathbf{k}} \psi_{\mathbf{k}}^{\dagger} - \beta_{\mathbf{k}} \psi_{\mathbf{k}+\mathbf{Q}}^{\dagger}.
\end{align}
For superpositions spanning all momentum sectors [Eq.~\eqref{eq:psi_general}], all $\phi_{1,2,3}$ components can be nonzero. By sampling randomized coefficients and then optimizing them, we obtain mean-field-like states whose order parameters $(\phi_1, \phi_2, \phi_3)$ trace a spherical manifold. However, this order parameter alone cannot confirm emergent $\mathrm{SU}(2)$ symmetry. In the following subsection, we analyze this through pseudospin operators derived from Brillouin zone folding.
 
\begin{figure*}[hbt!]
\begin{localgraphicspath}{{figs/draft_fig/}}
\includegraphics[width=0.6\columnwidth]{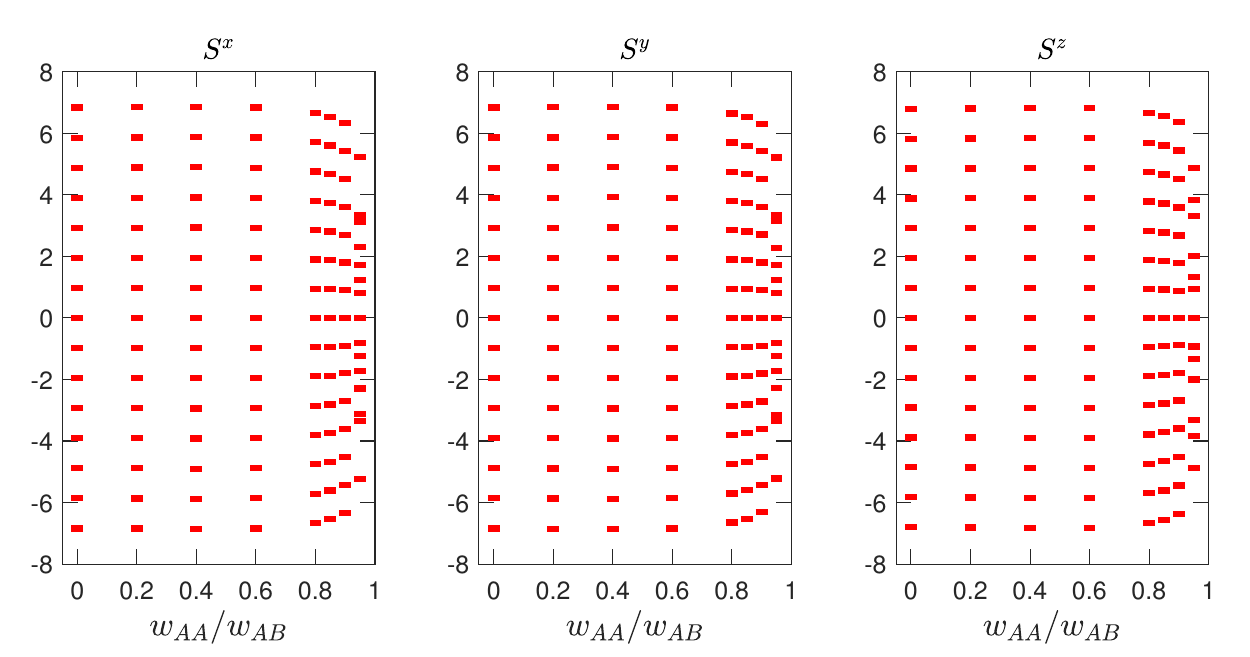}
\caption{$\hat{S}^{x/y/z}$ quantum numbers which are computed by diagonalizing the pseudo-spin $\hat{S}^{x/y/z}$ operators for $\nu_{\text{total}}=3+1/2$ on the $N_s=28$ cluster. The values are approximately $-N_{p}/2,-N_{p}/2+1,...,N_{p}/2-1,N_{p}/2$ for $w_{AA}/w_{AB}\le 0.9$ before the spectrum gap vanishes.} 
\label{fig:Sxyz}
\end{localgraphicspath}
\end{figure*}

\subsection{Emergent $\rm{SU}(2)$ symmetry}
The quasi-degenerate ground states and near-integer occupation spectra [Figs.~\ref{fig:correlation_matrix} (e)-(h)] provide strong evidence for emergent $\mathrm{SU}(2)$ ferromagnetism in dispersive Chern bands. Motivated by these observations and the Brillouin zone folding framework in Ref.~\onlinecite{dong2023many}, we define pseudospin operators through the following construction:
\begin{align}
\hat{S}^{x/y/z} & = \sum_{\mathbf{k} \in \text{folded BZ}} \hat{s}^{x/y/z}_{\mathbf{k}}, \notag\\
\hat{s}^x_{\mathbf{k}} & = \frac{1}{2} \left( \tilde{\phi}_{0,\mathbf{k}}^{\dagger} \tilde{\phi}_{1,\mathbf{k}} + \tilde{\phi}_{1,\mathbf{k}}^{\dagger} \tilde{\phi}_{0,\mathbf{k}} \right), \notag\\
\hat{s}^y_{\mathbf{k}} & = \frac{i}{2} \left( -\tilde{\phi}_{0,\mathbf{k}}^{\dagger} \tilde{\phi}_{1,\mathbf{k}} + \tilde{\phi}_{1,\mathbf{k}}^{\dagger} \tilde{\phi}_{0,\mathbf{k}} \right), \notag\\
\hat{s}^z_{\mathbf{k}} & = \frac{1}{2} \left( \tilde{\phi}_{0,\mathbf{k}}^{\dagger} \tilde{\phi}_{0,\mathbf{k}} - \tilde{\phi}_{1,\mathbf{k}}^{\dagger} \tilde{\phi}_{1,\mathbf{k}} \right),
\end{align}
where the momentum sum covers the folded Brillouin zone, and the orbital operators are defined as
\begin{align}
\tilde{\phi}_{0,\mathbf{k}}^{\dagger} & =\alpha_{\mathbf{k}}\psi_{\mathbf{k}}^{\dagger}+\beta_{\mathbf{k}}\psi_{\mathbf{k}+\frac{\mathbf{G}_{1}}{2}}^{\dagger},\notag\\
\tilde{\phi}_{1,\mathbf{k}}^{\dagger} & =\alpha_{\mathbf{k}+\frac{\mathbf{G}_{2}}{2}}\psi_{\mathbf{k}+\frac{\mathbf{G}_{2}}{2}}^{\dagger}- \beta_{\mathbf{k}+\frac{\mathbf{G}_{2}}{2}}\psi_{\mathbf{k}+\frac{\mathbf{G}_{1}}{2}+\frac{\mathbf{G}_{2}}{2}}^{\dagger}.
\end{align}
Here $\mathbf{G}_1/2=\mathbf{M}_2$, $\mathbf{G}_2/2=\mathbf{M}_1$, $\mathbf{G}_1/2+\mathbf{G}_2/2=\mathbf{M}_3$, and 
$\mathbf{G}_1,\mathbf{G}_2,\mathbf{G}_3$ form the reciprocal lattice basis of the triangular lattice. The coefficients $(\alpha_{\mathbf{k}}, \beta_{\mathbf{k}})$ are chosen to be the occupied orbitals of the $|+Z\rangle$ state ($\mathbf{Q} = \mathbf{M}_2$, order parameter $(0,\phi_2,0)$), while $(\alpha_{\mathbf{k}}, -\beta_{\mathbf{k}})$ describe the $|-Z\rangle$ state (order parameter $(0,-\phi_2,0)$). 
The $|\pm Z\rangle$ states constitute the north and south poles of the pseudospin Bloch sphere, establishing the quantization ($\hat{S}^z$) axis. Crucially, the orbitals $(\alpha_{\mathbf{k}}, \beta_{\mathbf{k}})$  depend not only on Bloch states, but also  on the Coulomb interaction, since they are inferred from linearly superposed many-body ground states. We observe $|\alpha_{\mathbf{k}}| \approx |\beta_{\mathbf{k}}|$ throughout the folded BZ, indicating nearly equal-weight superpositions of $\mathbf{k}$ and $\mathbf{k} + \mathbf{M}_2$ modes in the $|+Z\rangle$ state. This ensures approximate orthogonality $\langle +Z | -Z \rangle \approx 0$, consistent with opposite pseudospin orientations. With the pseudospin basis constructed, the generic ferromagnetic many-body state has the form
\begin{align}
\boldsymbol{\Phi}(\varphi,\phi)=\prod_{\mathbf{k}\in \text{folded BZ}}\left(\cos\frac{\varphi}{2} \tilde{\phi}_{0,\mathbf{k}}^{\dagger}+\sin\frac{\varphi}{2}e^{i\phi} \tilde{\phi}_{1,\mathbf{k}}^{\dagger}\right) |\text{Vac}\rangle,
\end{align}
where the Bloch sphere angle $(\theta,\phi)$ is linked to the CDW order parameters $(\phi_1,\phi_2,\phi_3)$ defined in Eq.~\eqref{eq:order}. 

For all $N_g$ ground states, we compute the $N_g \times N_g$ matrix representations of the pseudospin operators $\hat{S}^{x,y,z}$. A crucial $\mathrm{U}(1)$ gauge freedom exists in the  orbital definition: $\tilde{\phi}_{1,\mathbf{k}}^{\dagger} \rightarrow e^{i\Theta_{\mathbf{k}}} \tilde{\phi}_{1,\mathbf{k}}^{\dagger}$. While this transformation leaves $\hat{S}^z$ eigenvalues (as shown in the main text) unchanged, it affects $\hat{S}^x$ and $\hat{S}^y$. We fix this gauge by optimizing the phases $\Theta_{\mathbf{k}}$ to maximize the highest eigenvalue of $\hat{S}^x$, effectively aligning the pseudospin $x$-axis consistently across the folded Brillouin zone. Once the highest eigenvalue of $\hat{S}^x$ is optimized, highest eigenvalue of $\hat{S}^y$ is also optimized since there is no residual gauge degree of freedom in the pseudospin basis.

The resulting eigenvalues of $\hat{S}^{x,y,z}$ at various $w_{AA}$ are shown in Fig.~\ref{fig:Sxyz}. In the chiral limit ($w_{AA}=0$), these eigenvalues are approximately quantized and distributed between $-N_p/2$ and $N_p/2$. The slight deviation from exact quantization (e.g., maximum $S^z < N_p/2$) stems from imperfect orthogonality between the north pole ($|+Z\rangle$) and south pole ($|-Z\rangle$) orbital sets. 
Nevertheless, the near-integer quantization demonstrates approximate $\mathrm{SU}(2)$ pseudospin ferromagnetism with total spin $S = N_p/2$, where $N_p$ is the number of pseudospin sites (equal to the number of $\mathbf{k}$-points in the reduced Brillouin zone). This emergent symmetry persists away from the chiral limit, with comparable quantization quality until $w_{AA}/w_{AB} \approx 0.9$. Beyond this point ($w_{AA}/w_{AB} > 0.9$), the eigenvalues collapse into a continuous distribution, indicating destruction of ferromagnetic order. This abrupt change suggests a quantum phase transition near $w_{AA}/w_{AB} \approx 0.9$.

\begin{figure}[hbt!]
\begin{localgraphicspath}{{figs/draft_fig/}}
\includegraphics[width=0.5\columnwidth]{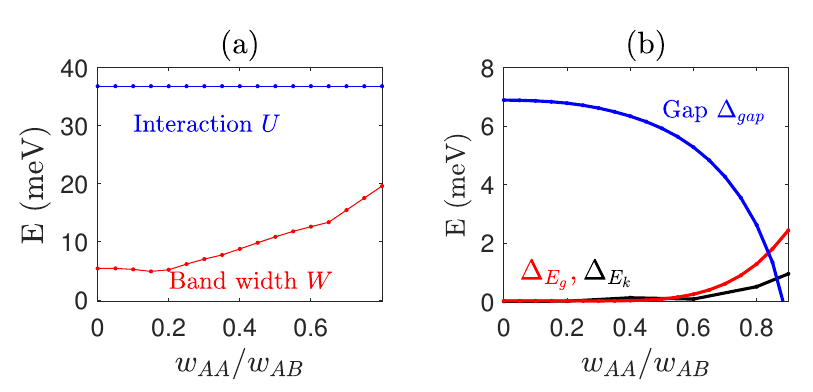}
\caption{Hamiltonian energy scales and ground state energy splitting  at $\nu_{\text{total}}=3+1/2$ on the $N_s=28$ cluster. (a) shows the bandwidth and typical Coulomb interaction strength. (b) shows the many-body spectrum gap $\Delta_{gap}$, the total energy splitting $\Delta_{E_g}$ between the $N_g$ ground states (red), and the kinetic energy $H_0=\sum_{k}\epsilon_k\hat{n}_k$ splitting $\Delta_{E_k}$ between the $N_g$ ground states (black).  }
\label{fig:spectrum_split}
\end{localgraphicspath}
\end{figure}

\subsection{Energy splitting of ground states}

Aside from pseudospin ferromagnetism, the emergent $\mathrm{SU}(2)$ symmetry manifests through suppressed ground-state energy splitting. Figure~\ref{fig:spectrum_split}(a) shows the characteristic Coulomb energy scale $U$ and bandwidth $W$ at total filling $\nu_\mathrm{total} = 3 + 1/2$. The bandwidth increases with $w_{AA}/w_{AB}$, reaching $W \approx 15$ meV at $w_{AA}/w_{AB} = 0.6$. We estimate the Coulomb scale as $U = V(|\mathbf{G}_1|)N_s/(2A) \approx 40$ meV (for dielectric constant $\epsilon = 4$), where $A/N_s$ is the moiré unit cell area and $|\mathbf{G}_1|$ is the reciprocal lattice vector magnitude.
Figure~\ref{fig:spectrum_split}(b) displays three key energy scales: (i) many-body excitation gap $\Delta_{\mathrm{gap}} = E_{N_g} - E_{N_g-1}$,
(ii) ground-state splitting $\Delta_{E_g} = E_{N_g-1} - E_0$ (energy spread within the $N_g$-fold manifold), and (iii) kinetic energy variance $\Delta_{E_k}$ ($H_0 = \sum_{\mathbf{k}} \epsilon_{\mathbf{k}} \hat{n}_{\mathbf{k}}$ spread).
The gap $\Delta_{\mathrm{gap}}$ closes near $w_{AA}/w_{AB} \approx 0.9$, consistent with the critical point identified from pseudospin quantization (Fig.~\ref{fig:Sxyz}). Remarkably, for $w_{AA}/w_{AB} < 0.6$, $\Delta_{E_g}$ remains strongly suppressed ($\sim 0.1$ meV) compared to the bandwidth $W \sim 5-15$ meV. Similarly, $\Delta_{E_k}$ is suppressed to a similar magnitude, evidenced by nearly identical momentum distributions across the $N_g$ ground states [Figs.~\ref{fig:correlation_matrix}(a)-(d)]. 
This suppression of energy splitting demonstrates that the $\mathrm{SU}(2)$ symmetry is surprisingly robust and approximately manifests across a significant parameter range away from the chiral limit.  



\section{Further numerical analysis for the fractional Chern insulator }

\begin{figure}[hbt!]
\begin{localgraphicspath}{{figs/draft_fig/}}
\includegraphics[width=0.4\columnwidth]{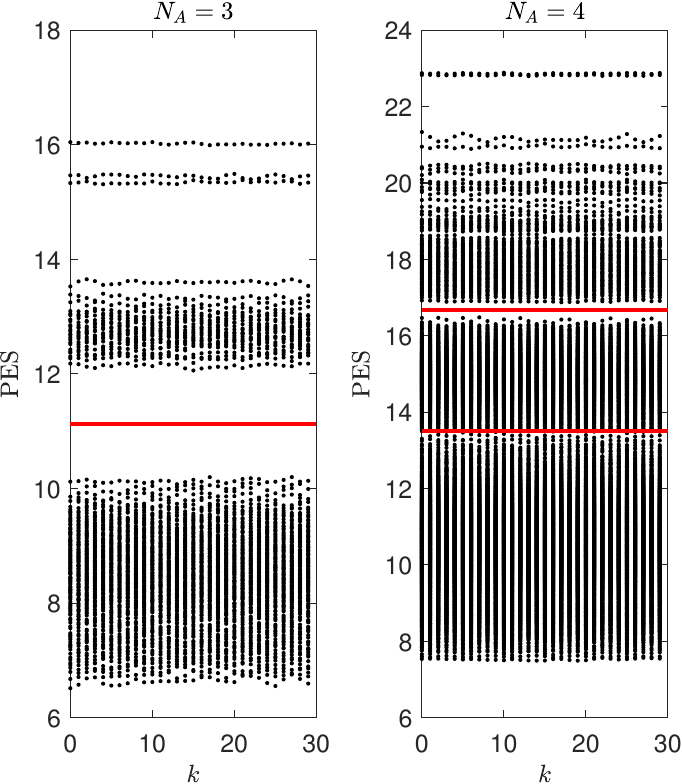}
\caption{Particle entanglement spectrum (PES) for $\nu_{\text{total}}=3+1/3, N_s=30, w_{AA}=0.3w_{AB}$ with different subsystem particle number $N_A=3$ and $4$. For $N_A=3$, the number of states below the PES gap is 3250. For $N_A=3$, the number of states below the two PES gaps are 17250 and 24840. }
\label{fig:PES}
\end{localgraphicspath}
\end{figure}

To identify the topological order of the fractional Chern insulator at $\nu=1/3$, we compute the particle entanglement spectrum (PES)~\cite{regnault2011fractional} which is defined by eigenvalues of the reduced density matrix in the orbital occupation number basis. The level counting of the PES reflects a generalized Pauli principle that provides a fingerprint for distinguishing different fractional quantum Hall states. For $w_{AA}=0.3w_{AB}$, we show the PES with subsystem particle numbers $N_A=3$ and $4$ in Fig.~\ref{fig:PES}. The number of states below the PES gaps that separate the low- and higher-energy excitations all match with the Halperin (112) state~\cite{dong2023many,liu2024engineering}. For $w_{AA}=0.6$, we find that the PES gap is only clear for $N_A=3$. Nevertheless, we believe that $w_{AA}=0.6$ and $w_{AA}=0.3$ belong to the same phase for $\nu_{\text{total}}=3+1/3$ since the many-body spectrum gap is nearly constant in this parameter range.

\section{pair correlation functions}
\begin{figure}[hbt!]
\begin{localgraphicspath}{{figs/draft_fig/}}
\includegraphics[width=\columnwidth]{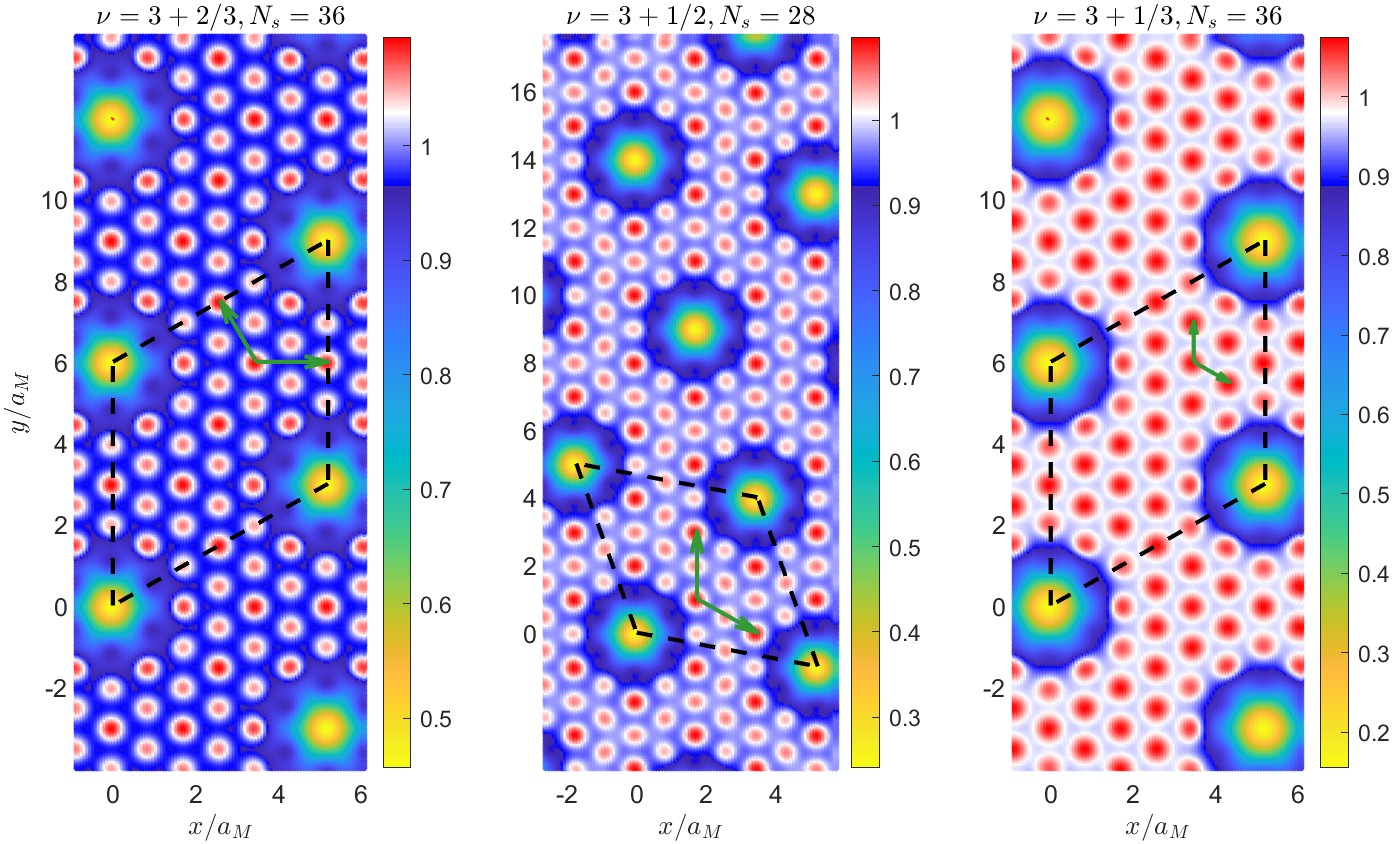}
\caption{Real space pair correlation functions $g(r-r')$ for $w_{AA}/w_{AB}=0.6$ at three filling fractions. The black dashed lines marks the periodic real space clusters of size $N_s$. The lattice vectors of each state are marked by green arrows. }
\label{fig:gr}
\end{localgraphicspath}
\end{figure}

In the main text only momentum space structure factors are shown. Here we provide further real space pair correlation functions to further confirm symmetry broken/unbroken orders of different phases. The real space pair correlation function is defined as~\cite{reddy2023toward}
\begin{align}
g(\mathbf{r}-\mathbf{r}')=\frac{1}{N_p^2}\sum_{\mathbf{q}} \langle :\tilde{\rho}_{\mathbf{q}} \tilde{\rho}_{-\mathbf{q}}:\rangle e^{i\mathbf{q}(\mathbf{r}-\mathbf{r}')}.
\label{eq:gr}
\end{align}
The values of $g(r-r')$ in Fig.~\ref{fig:gr}---for three different fillings---form a periodic pattern with period $N_s$ due to finite discretization of the momentum grid. The yellow circles at $r\approx r'$ correspond to minima, reflecting the repulsive nature of the Coulomb interaction. Away from $r\approx r'$, uniform crystal structures can be recognized with unit cell areas $\sqrt{3}a_{M}\times\sqrt{3}a_{M}$, $2a_{M}\times 2a_{M}$, and $a_{M}\times a_{M}$ in the three cases, respectively. The elementary lattice vectors of each state are labeled by green arrows. For the quantum Hall ferromagnet at $\nu_{\rm{total}}=3+1/2$, we clarify that the $2\times 2$ enlarged moir\'e cell is a consequence of averaging over extensive ground states, as we have seen that different linear combinations from the degenerate ground states yield different ordering wavevectors $\mathbf{Q}$.

\bibliography{draft.bib}